\documentclass[conference,a4paper]{APSIPA2025}

\usepackage{amsmath,amsfonts}
\usepackage{graphicx}
\usepackage{multirow}
\usepackage{threeparttable}
\usepackage{booktabs}
\usepackage{comment}
\usepackage{xcolor}     % 提供 \textcolor
\usepackage{url}        % 提供 \url
\usepackage[numbers,sort&compress]{natbib}  % 走 BibTeX 路线

\usepackage{geometry}
\usepackage{fancyhdr}

\fancypagestyle{firststyle}{
  \fancyhf{}
  \fancyhead[C]{}
}

\begin{document}
\title{Reference-free Adversarial Sex Obfuscation in Speech}

\author{
\authorblockN{
Yangyang Qu\authorrefmark{1}, Michele Panariello\authorrefmark{1},  Massimiliano Todisco\authorrefmark{1} and
Nicholas Evans\authorrefmark{1}
}

\authorblockA{
\authorrefmark{1}
EURECOM, France \\
 }

}

\maketitle
\thispagestyle{firststyle}
\pagestyle{fancy}

\begin{abstract}
  Sex conversion in speech involves privacy risks from data collection and often leaves residual sex-specific cues in outputs, even when target speaker references are unavailable. We introduce RASO for Reference-free Adversarial Sex Obfuscation. Innovations include a sex-conditional adversarial learning framework to disentangle linguistic content from sex-related acoustic markers and explicit regularisation to align fundamental frequency distributions and formant trajectories with sex-neutral characteristics learned from sex-balanced training data.  RASO preserves linguistic content and, even when assessed under a semi-informed attack model, it significantly outperforms a competing approach to sex obfuscation.
\end{abstract}

\section{Introduction}  
Voice Conversion (VC) plays a critical role in privacy-sensitive applications, e.g.\ anonymisation~\cite{tomashenko2024voiceprivacy} of speech data collected in healthcare scenarios. 
Privacy preservation involves the obfuscation of speaker-specific traits (e.g., the voice, sex, age and accent) but the preservation of utility (e.g., the linguistic content, naturalness, prosody, emotion and health-related cues). The work presented in this paper concerns obfuscation of the speaker's sex.\footnote{As in~\cite{noe21_interspeech}, sex refers to biological attributes, whereas gender refers to socially constructed roles
and behaviour~\cite{prince2005sex}.}  Traditional voice conversion methods, which rely on parallel corpora or target speaker references~\cite{kaneko2018cyclegan,kameoka2018stargan}, face two fundamental limitations, namely the high cost of acquiring sensitive target speech data and the incomplete suppression of sex-discriminative acoustic features (e.g., fundamental frequency distributions, formant trajectories) in zero-shot scenarios, which leaves residual cues exploitable by re-identification attacks~\cite{noe2023hiding}.  

To address these challenges, we propose RASO, a GAN-based framework for reference-free, sex-neutral voice conversion. Our approach introduces the following key innovations: 

1. Reference-free, sex-neutral conversion via conditional adversarial learning. Our learning framework disentangles speaker-agnostic linguistic content from sex-discriminative acoustic features (fundamental frequency(F0) distributions and formant trajectories). A discriminator enforces sex ambiguity in generated speech, enabling the obfuscation of sex-specific attributes without requiring reference target speaker data.  

2. Explicit acoustic regularisation for distributional neutrality. To ensure sex neutrality, we introduce a sex feature modification module that normalises the probability density of fundamental frequency distributions and the temporal dynamic range of formant trajectories to align with mixed-sex speech statistics. This mechanism eliminates sex-specific offsets in acoustic parameters to achieve population-level, sex-neutral acoustic representations.

By integrating these mechanisms, RASO offers a robust solution which eliminates the need for sensitive target speaker data, effectively suppresses sex-related attributes while maintaining high speech intelligibility and naturalness, and ensures population-level privacy by aligning acoustic features with mixed-sex statistical distributions. Experimental results show that RASO surpasses competing state-of-the-art methods~\cite{noe2023hiding,panariello2024speaker}.

\section{Related Work}

Deep learning has propelled voice conversion advancements, with GAN-based methods like CycleGAN-VC~\cite{kaneko2018cyclegan} and StarGANv2-VC~\cite{li2021starganv2} leading the field by disentangling linguistic content from speaker attributes via cycle-consistency or style encoding for non-parallel, multi-domain conversion. These models excel at generating high-fidelity prosodic details, such as pitch contours, rhythm, and timbral nuances, but inadvertently retain privacy-sensitive speaker cues (e.g., sex-specific formant patterns, vocal tract characteristics) embedded in their representations, as their design prioritises identity preservation over attribute obfuscation. 
%Variational autoencoders like VQMIVC~\cite{wang2021vqmivc} leverage vector quantisation to separate content and speaker factors, yet lack explicit mechanisms to suppress sensitive features, while diffusion models like DiffGAN-VC~\cite{zhang2023voice} improve conversion quality through iterative denoising of spectrograms but focus on reconstruction accuracy rather than privacy-protective feature suppression. 

In the realm of speaker anonymisation, recent advancements have sought to balance privacy preservation with linguistic utility. Early work by Fang et al.~\cite{fang2019speaker} introduced a foundational approach by fusing speaker X-vectors with neural waveform models, enabling identity obfuscation while retaining linguistic content. Building on this, Srivastava et al.~\cite{srivastava2020design} refined the approach by introducing pseudo-speaker selection strategies that dynamically mixed X-vectors to enhance privacy-utility trade-offs. Later, Champion~\cite{champion2023anonymizing} proposed quantisation-based transformations to suppress speaker-related information in acoustic features, outperforming traditional noise-based methods. Concurrently, Panariello et al.~\cite{panariello2024speaker} adopted a neural audio codec strategy, leveraging pre-trained EnCodec and Transformer architectures to disentangle semantic-acoustic tokens for synthesis. Meyer et al.~\cite{meyer2023prosody} further advanced the field by generating pseudo-embeddings via GANs to replace speaker identities while preserving prosodic nuances. Tomashenko et al.~\cite{tomashenko2024voiceprivacy} evaluated these systems under semi-informed attacker models, emphasising the need for standardised frameworks to assess multi-condition privacy-utility trade-offs.

There is less work in sex obfuscation. Stoidis and Cavallaro~\cite{stoidis2022generating} introduced GenGAN, which generates sex-ambiguous speech by smoothing spectral differences, achieving balanced privacy and speech intelligibility.  Noé et al.~\cite{noe2023hiding} propose a "zero-evidence" framework using adversarial training and normalising flows to suppress sex information in an analysis/synthesis pipeline. Chouchane et al.~\cite {chouchane2023differentially} present a differentially private adversarial auto-encoder framework,  designed to protect sex information in voice biometrics by mitigating sex-specific cues. In their other work~\cite {chouchane2023fairness}, they analysed how sex affects voice biometric systems and proposed strategies to reduce sex-related biases. Koutsogiannaki et al.~\cite{koutsogiannaki2024gender} propose a method that blends low-frequency spectral characteristics with prosodic patterns to generate sex-ambiguous speech outputs and reduce the discriminability of sex attributes in speech signals. 

\begin{figure*}[!htb]
    \centering 
    \includegraphics[width=0.80\textwidth]{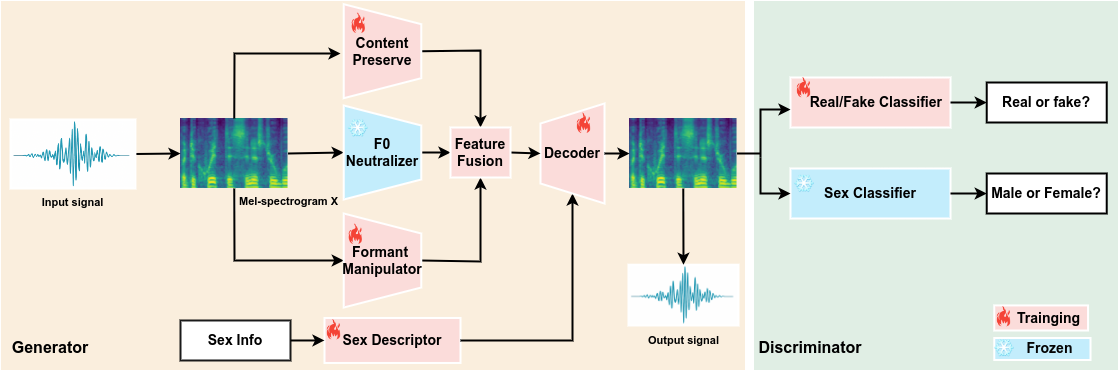} 
    \caption{Architecture of the RASO framework. The left side shows the training process of the generator. The right side shows the training process of the discriminator.}
    \label{fig:model_architecture}
\end{figure*}

\section{Model Architecture}  
To achieve sex obfuscation in speech, our proposed framework employs a privacy-driven adversarial architecture that suppresses sex-discriminative acoustic features while preserving linguistic content. The model consists of two core components:  a generator for feature-level de-identification and a multi-task discriminator. The architecture is illustrated in Fig.~\ref{fig:model_architecture} and is described in the following.

\subsection{Generator: sex feature suppression network}  
The generator aims to remove sex-specific acoustic markers from input speech while keeping other information intact. It employs a dual-branch architecture which explicitly decouples linguistic content from sex feature suppression.

\subsubsection{Linguistic content preservation}  
A Mel-spectrogram encoder is employed to extract linguistic information. The input Mel-spectrogram $\mathbf{X} \in \mathbb{R}^{B \times 1 \times 80 \times T}$ is compressed into a latent content vector by a hierarchical encoding module consisting of residual blocks with downsampling. The input Mel-spectrogram $\mathbf{X} \in \mathbb{R}^{B \times 1 \times 80 \times T}$ is compressed into a latent content vector $\mathbf{Z}_{\text{cont}} \in \mathbb{R}^{B \times C \times H \times W}$ by a hierarchical encoding module consisting of residual blocks with downsampling, where $C$ denotes the channel dimension and $H, W$ represent spatial dimensions. 

\subsubsection{Sex feature modification}  
Three specialised modules are employed to neutralise sex-discriminative acoustic features and preserve semantic content:

\textbf{Formant Manipulation Branch -}  This module processes the lower 40 Mel bands ($\mathbf{F}_{\text{low}} \in \mathbb{R}^{B \times 1 \times 40 \times T}$) to suppress sex-discriminative formant patterns. By introducing a sex-conditioned embedding mechanism, the formants for each sex are edited according to:
\[  
\mathbf{X}_{\text{mod}} = \mathbf{X} \odot \left(1 + \mathbf{W} \cdot \mathbf{s}(\mathbf{y}_{\text{org}})\right)  
\]  
where $\mathbf{W} \in \mathbb{R}^{40 \times 64}$ is a learnable projection matrix, and $\mathbf{s}(\mathbf{y}_{\text{org}}) \in \mathbb{R}^{64}$ is an embedding vector generated by the sex descriptor based on the input sex label $\mathbf{y}_{\text{org}} \in \{0, 1\}$ ($0$ represents male, $1$ represents female). Label $\mathbf{s}(\mathbf{y}_{\text{org}})$ is parameterised independently via a conditional embedding layer, enabling the module to apply sex-specific frequency modulation strategies.  

For female inputs ($\mathbf{y}_{\text{org}} = 1$), $\mathbf{s}(0)$ is optimised to enhance attenuation at higher frequencies to neutralise female-specific formant concentrations. Conversely, for male inputs ($\mathbf{y}_{\text{org}} = 0$), $\mathbf{s}(1)$ targets low-frequency bands to suppress male-dominant spectral features.  During training, $\mathbf{s}(\mathbf{y}_{\text{org}})$ and $\mathbf{W}$ are jointly optimised. This design eliminates the need for target speaker references, relying solely on binary sex labels to achieve directional suppression, which effectively obfuscates sex-related acoustic cues while preserving linguistic content.

\textbf{F0 Neutralization Branch -}  
The fundamental frequency contour \( f_0^{\text{pred}} \) is predicted by the model JDC~\cite{kum2019joint}. 
\footnote{\url{https://github.com/keums/melodyExtraction_JDC}} The predicted fundamental frequency contour \( f_0^{\text{pred}} \) is mapped to the sex-neutralized counterpart \( f_0^{\text{shifted}} \) via log-domain shifting:
% This is normalized to a dynamic baseline $\mu_{\text{neu}}$ through log-domain shifting:  
\begin{equation}  
f_0^{\text{shifted}} = \exp\left(\log(f_0^{\text{pred}}) + \log\left(\frac{\mu_{\text{neutral}}}{\bar{f}_0^{\text{org}}}\right)\right)  
\end{equation}  
where $\bar{f}_0^{\text{org}}$denotes the global mean F0 of the input speech and where $\mu_{\text{neutral}}$ is updated via exponential moving average during training:  
\begin{equation}  
\mu_{\text{neutral}}^{(t)} = \gamma \mu_{\text{neutral}}^{(t-1)} + (1-\gamma) \cdot \bar{f}_0^{\text{batch}} 
\end{equation}  
where $\gamma=0.99$ and $\bar{f}_0^{\text{batch}}$ represent the average F0 across all speech samples in the current training batch, which acts to ensure that $\mu_{\text{neutral}}$ dynamically approximates the global F0 statistics of the mixed-sex training corpus while suppressing batch-specific fluctuations.

\textbf{Feature Fusion and Reconstruction Module -}  
The outputs of the sex feature suppression branch, including formant-suppressed low-frequency Mel-bands and F0-neutralised contours, are fused with the content representation $\mathbf{Z}_{\text{cont}}$ via a formant-guided attention mechanism which extracts sex-relevant spectral patterns from the lower 40 Mel-bands and generates attention maps through style embeddings to highlight sex-neutral frequency regions. Fused features are processed through upsampling residual blocks with adaptive instance normalisation (AdaIN)\cite{huang2017arbitrary} to restore spectral resolution, followed by a projection layer to reconstruct the Mel-spectrogram. This design suppresses sex-discriminative acoustic cues (formant shifts, F0 trends) while preserving linguistic content through multi-scale feature refinement, enabling reference-free sex obfuscation with high-fidelity speech synthesis.

\subsection{Discriminator: Adversarial Privacy Transformation}  

The discriminator \( D \) is designed with a dual-objective architecture to enforce two complementary objectives in the adversarial training framework:  speech generation with preserved speech intelligibility and effective sex neutrality. 

\subsubsection{Real/Fake Discrimination}
A multi-scale convolutional network with spectral normalisation is used to distinguish between real spectrograms $\mathbf{X}_{\text{real}}$ and generated spectrograms $\hat{\mathbf{X}}$. A least-squares generative adversarial loss~\cite{mao2017least} is used to stabilise training.  

\subsubsection{Sex Confusion Discrimination}  
A pre-trained sex classifier with frozen parameters is used to evaluate the sex ambiguity of generated speech~\cite{burkhardt2023speech}.\footnote{\url{https://huggingface.co/audeering/wav2vec2-large-robust-24-ft-age-gender}} During training, the discriminator provides gradients to the generator to maximise the classifier output entropy over $\hat{\mathbf{X}}$, while the classifier parameters remain fixed to provide an unbiased evaluation.  

%Unlike sex conversion (male-to-female or female-to-male), the objective is to render speech statistically sex-ambiguous or sex neutral.

\subsection{Loss Functions}  
Our privacy-driven loss framework balances sex obfuscation and speech intelligibility through a multi-objective optimisation strategy.  This is achieved using a set of loss functions, each of which is described below.  

\subsubsection{Adversarial Loss}  
We adopt the Least Squares GAN (LSGAN) loss \cite{mao2017least} with soft labels to stabilise training and promote spectrogram results\cite{salimans2016improved}:  
\begin{align}
\mathcal{L}_D^{\text{adv}} = 
&\frac{1}{2} \mathbb{E}_{\mathbf{X}_{\text{real}} \sim p_{\text{data}}} \left[ \left(D(\mathbf{X}_{\text{real}}) - 0.95\right)^2 \right] \nonumber \\
&+ \frac{1}{2} \mathbb{E}_{\hat{\mathbf{X}} \sim p_{\text{gen}}} \left[ \left(D(\hat{\mathbf{X}}) - 0.05\right)^2 \right],
\label{eq:discriminator_adv_loss}
\end{align}
where $\mathbf{X}_{\text{real}}$ denotes real speech Mel-spectrograms, $\hat{\mathbf{X}}$ denotes generated sex-neutral speech, and $D$ denotes the discriminator. The soft labels ($0.95$ for real, $0.05$ for fake) mitigate gradient vanishing compared to hard labels ($1$ and $0$).   The adversarial loss of the generator is given by:  
\begin{equation}  
\mathcal{L}_G^{\text{adv}} = \frac{1}{2} \mathbb{E}_{\hat{\mathbf{X}} \sim p_{\text{gen}}} \left[ \left(D(\hat{\mathbf{X}}) - 0.95\right)^2 \right].  
\label{eq:generator_adv_loss}  
\end{equation}

\subsubsection{Sex Ambiguity Loss}  
To enforce sex neutrality, we maximise the entropy of a pre-trained sex classifier $C$ \cite{burkhardt2023speech} over generated speech. The loss is defined as the negative Shannon entropy:  
\begin{align}  
\mathcal{L}_{\text{sex}} = - \mathbb{E} \bigg[ 
    &\mathcal{P}_{\text{male}}(\hat{\mathbf{X}}) \cdot \log \mathcal{P}_{\text{male}}(\hat{\mathbf{X}}) \nonumber \\
    &+ \left(1 - \mathcal{P}_{\text{male}}(\hat{\mathbf{X}})\right) \cdot \log \left(1 - \mathcal{P}_{\text{male}}(\hat{\mathbf{X}})\right) 
\bigg],  
\label{eq:sex_entropy_loss}  
\end{align}   
where $\mathcal{P}_{\text{male}}(\hat{\mathbf{X}}) \in [0, 1]$ is the probability that the outcome is classified as male. Minimizing $\mathcal{L}_{\text{sex}}$ forces $\mathcal{P}_{\text{male}} \to 0.5$, ensuring a uniform class distribution.  

\subsubsection{Content Preservation Loss}  
To ensure linguistic content is retained during transformation, we employ a feature-level consistency loss using a pre-trained automatic speech recognition (ASR) model \cite{kim2017joint}.\footnote{\url{https://github.com/yl4579/AuxiliaryASR}} The loss is defined as:  
\begin{equation}  
\mathcal{L}_{\text{content}} = \mathbb{E}_{\mathbf{X}} \left[ \left\lVert h_{\text{asr}}(\mathbf{X}) - h_{\text{asr}}(\hat{\mathbf{X}}) \right\rVert_1 \right],  
\label{eq:content_loss}  
\end{equation}  
where \( h_{\text{asr}}(\cdot) \) denotes the contextual feature extractor from an ASR model encoder—a network which captures phonetic and semantic dependencies in speech. Here, \( \mathbf{X} \) represents the original speech signal, \( \hat{\mathbf{X}} \) is the transformed output, and \( \lVert \cdot \rVert_1 \) is the L1 norm, which minimizes the absolute difference between high-level features of the original and generated speech.

\subsubsection{Cycle Consistency Loss}  
To mitigate content degradation during sex obfuscation, a cycle consistency loss is introduced to enforce bidirectional fidelity between the original and transformed speech. The loss is defined as:  
\begin{equation}  
\mathcal{L}_{\text{cyc}} = \mathbb{E}_{\mathbf{X}, s_{\text{src}}} \left[ \left\lVert G\left(G(\mathbf{X}, s_{\text{neutral}}), s_{\text{src}}\right) - \mathbf{X} \right\rVert_1 \right],  
\label{eq:cycle_loss}  
\end{equation}  
% where \( s_{\text{src}} = E_g(\mathbf{X}) \) denotes the gender embedding extracted by the sex descriptor branch. The term $s_{\text{neutral}}$  represents a vector that serves as the target for gender neutralization, prompting the generator \( G \) to produce speech with obfuscated gender cues. The nested application of \( G \) first transforms the input speech \( \mathbf{X} \) into a gender-neutral representation using \( s_{\text{neutral}} \), then reconstructs the original speech by reintroducing the source gender embedding \( s_{\text{src}} \). By minimizing the L1 distance between the reconstructed and original spectrograms, this mechanism ensures that the obfuscation-reconstruction process is invertible, thereby preserving linguistic content while neutralizing gender-specific acoustics.
where \(s_{\text{src}}\) is the source sex embedding extracted by the sex descriptor branch, and \(s_{\text{neutral}}\) is the sex-neutral target vector. By minimising the L1 distance between reconstructed and original spectrograms, this mechanism forces the generator G to learn an invertible mapping, preserving linguistic content while neutralising sex-specific acoustics.

\subsubsection{F0 Neutralization Loss}  
The F0 is normalised to a dynamic neutral baseline $\mu_{\text{neu}}$ (initialised at 150 Hz, the median F0 of mixed-sex training data) while preserving relative pitch dynamics:  
\begin{align}  
\mathcal{L}_{\text{F0}} = \mathbb{E} \bigg[ 
    &\left\lVert \bar{f}_0^{\text{gen}} - \mu_{\text{neu}} \right\rVert_1 \nonumber \\
    &+ \lambda_{\text{rel}} \cdot \left\lVert \Delta\log(f_0^{\text{gen}}) - \Delta\log(f_0^{\text{org}}) \right\rVert_1 
\bigg],  
\label{eq:f0_loss}  
\end{align}  
where $\bar{f}_0^{\text{gen}}$ and $\bar{f}_0^{\text{org}}$ denote the mean F0 values of generated and original speech, respectively; $\Delta\log(f_0) = \log(f_0) - \log(\bar{f}_0)$ represents log-normalized pitch contours capturing relative dynamics; and $\lambda_{\text{rel}} = 0.8$ balances absolute F0 alignment and relative pitch preservation.

\subsubsection{Formant Suppression Loss}  
Generated formants are aligned with mixed-sex statistical moments (mean $\mu$ and standard deviation $\sigma$):  
\begin{align}  
\mathcal{L}_{\text{formant}} = \sum_{k=1}^{3} \bigg( 
    &\left\lVert \mu(\mathbf{F}_k^{\text{gen}}) - \mu(\mathbf{F}_k^{\text{neutral}}) \right\rVert_1 \nonumber \\
    &+ \beta \cdot \left\lVert \sigma(\mathbf{F}_k^{\text{gen}}) - \sigma(\mathbf{F}_k^{\text{neutral}}) \right\rVert_1 
\bigg),  
\label{eq:formant_loss}  
\end{align}   
where: $\mathbf{F}_k^{\text{gen}}$ denotes the $k$-th formant of generated speech (extracted via Linear Predictive Coding); $\mu(\mathbf{F}_k^{\text{neutral}})$ and $\sigma(\mathbf{F}_k^{\text{neutral}})$ are the global mean and standard deviation of formants computed from mixed-sex training data; $\beta = 0.3$ controls formant smoothness to balance neutrality and naturalness.

\subsubsection{Total Generator Loss}  
The total generator loss is used with empirically tuned weights to balance sex obfuscation and speech intelligibility:  
\begin{align}  
\mathcal{L}_G = 
&\alpha_1 \mathcal{L}_G^{\text{adv}} + \alpha_2 \mathcal{L}_{\text{sex}} + \alpha_3 \mathcal{L}_{\text{content}} \nonumber \\
&\alpha_4 \mathcal{L}_{\text{F0}} + \alpha_5 \mathcal{L}_{\text{formant}} + \alpha_6 \mathcal{L}_{\text{cyc}},  
\label{eq:total_generator_loss}  
\end{align}  
with weights determined from a grid search using a validation set: $\alpha_1=1.0$, $\alpha_2=5.0$ (prioritising sex neutrality), $\alpha_3=10.0$ (critical for content preservation), $\alpha_4=2.0$, $\alpha_5=1.0$, and $\alpha_6=10.0$.

\section{Experiments}
 
\subsection{Dataset}
Inspired by previous research on voice privacy\cite{tomashenko2024voiceprivacy}, we employ the LibriSpeech corpus~\cite{panayotov2015librispeech} for experiments, specifically the train-clean-360 subset for training and test-clean subset for evaluation. The training set contains speech from 921 speakers (482 male, 439 female), while the test set includes 40 unseen speakers (20 male, 20 female). The large-scale, high-quality recordings in the train-clean-360 subset ensure robust model training, while the test-clean subset provides a controlled, unseen dataset for rigorous privacy and conversion quality assessment.

\begin{table}[t!]
\centering
\begin{threeparttable}
\caption{Performance Comparison Under Different Attacker Scenarios\\
(EER$\uparrow$ indicates higher sex classification error for better privacy; WER$\downarrow$ indicates lower speech recognition error for better intelligibility)}
\label{combined-eer-wer-results}
\begin{tabular}{lcc c}
\toprule
\multirow{2}{*}{\textbf{Model Type}} & \multicolumn{2}{c}{\textbf{Ignorant Attacker}} & \textbf{Semi Informed} \\
\cmidrule(r){2-3} \cmidrule(l){4-4}
& \textbf{EER (\%) $\uparrow$} & \textbf{WER (\%) $\downarrow$} & \textbf{EER (\%) $\uparrow$} \\
\midrule
Raw Data                  & 7.22      & 1.84    & -- \\
Pan. et al. \cite{panariello2024speaker}      & 48.56     & 5.90     & 32.15 \\
Noe et al. \cite{noe2023hiding}          & 36.88      & 2.48     & 16.37 \\
RASO                                       & \textbf{55.38}     & \textbf{2.47}    & \textbf{47.25} \\
\bottomrule
\end{tabular}
\end{threeparttable}
\end{table}

\subsection{Training Details}  
We employ the AdamW optimizer~\cite{loshchilov2017decoupled} with learning rates of \(10^{-5}\) for the generator and \(10^{-4}\) for the discriminator.  Training is performed with a batch size of 64 and an NVIDIA 3090 GPU with PyTorch mixed-precision acceleration. Early stopping based on the validation loss is applied for 150 epochs.

\subsection{Objective Metrics} 
We adopt the Equal Error Rate (EER) to evaluate sex classification and the Word Error Rate (WER) to evaluation ASR performance. The EER is derived from the pre-trained sex classifier~\cite{burkhardt2023speech} and quantifies the obfuscation of sex-specific acoustic features, while the WER relies on a pre-trained ASR system~\cite{speechbrain} trained on the full LibriSpeech-train-960 dataset to assess the preservation of linguistic content.

\subsection{Evaluations}
Within the context of voice privacy protection, our evaluation of RASO incorporates two state-of-the-art baselines, each contextualised by their relationship to sex obfuscation. Noe et al. \cite{noe2023hiding}, explicitly designed for sex obfuscation, serves as a direct comparator. Complementing this, Panariello et al.~\cite{panariello2024speaker} is included to benchmark a related approach. Although their work focuses on speaker anonymisation, it hides the speakerID while also hiding sex-related features.

To simulate adversarial scenarios of increasing sophistication, we adopt two attack models inspired by the VoicePrivacy Challenge~\cite{tomashenko2024voiceprivacy}. The first, an ignorant attack model, assumes the attacker lacks knowledge of RASO and uses a pre-trained sex classifier \footnote{\url{https://huggingface.co/audeering/wav2vec2-large-robust-24-ft-age-gender}} to classify obfuscated speech. In the second scenario, a semi-informed attack~\cite{srivastava2020evaluating}, the attacker fine-tunes a sex classifier on sex-neutralised datasets generated by Noe et al.~\cite{noe2023hiding}, Panariello et al.~ \cite{panariello2024speaker} and RASO, respectively. This setup assesses RASO’s resilience against classifiers adapted to obfuscation patterns from competing methods, providing a rigorous comparison 
%of privacy-preservation robustness 
across frameworks.

% \subsection{Result Analysis}  
Results for our system and two competing methods are presented in Table~\ref{combined-eer-wer-results} for both attack models.
%presents a comparative analysis of sex obfuscation performance under two attacker scenarios: ignorant attackers and semi-informed attackers. The metrics include EER for sex classification and WER for speech intelligibility. 
%The results compare the proposed RASO framework against state-of-the-art baselines (Panariello et al.\cite{{panariello2024speaker}} and Noe et al.\cite{noe2023hiding}) and
Also shown are results for raw (unprocessed/unprotected) speech data.  
For the ignorant attack model, RASO achieves an EER of 55.38\%, significantly outperforming results for both competing systems -- 36.88\% for Noe et al.~\cite{noe2023hiding} and 48.56\% for Panariello et al.~\cite{panariello2024speaker}.  The latter result shows that even voice anonymisation systems, though not tuned specifically for sex obfuscation, can still be effective, most likely because target/pseudo speaker voices used in the conversion are of random sex. RASO maintains a WER of 2.47\%, comparable to that of Noe et al.\ (2.48\%) but far superior to that of Panariello et al.\ (5.90\%).  Together, these results demonstrate the successful suppression of sex-specific acoustic features (e.g., formant patterns, F0 contours) and the preservation of linguistic content. 

The results for the semi-informed attack model exhibit even more pronounced disparities, underscoring the efficacy of our approach. RASO achieves an EER of 47.25\%, significantly outperforming the 32.15\% and 16.37\% for competing systems. 
%proposed by Panariello et al.~\cite{panariello2024speaker} and Noe et al.~\cite{noe2023hiding}, respectively. 
This substantial improvement highlights the resilience conferred by adversarial training and our multi-task loss design against more sophisticated attacks, still without access to target speaker data. Across both attack models, RASO consistently attains a high EER and low WER.
%, thus establishing a superior equilibrium between privacy preservation and speech intelligibility in reference-free sex obfuscation.

\section{Conclusions}
We propose an integrated adversarial framework 
for robust sex obfuscation without target speaker references. Our approach adjusts formant patterns and F0 distributions to neutralise sex cues in speech while preserving intelligibility. Experimental results confirm improvements over competing methods, demonstrating the merit of our approach in balancing the obfuscation of sex information with the preservation of linguistic content.

In future research, a potential extension could involve introducing mechanisms to control the degree of sex obfuscation, which would allow users to tailor the conversion intensity according to specific privacy requirements, thereby enhancing the framework’s adaptability across diverse application domains.

% \section*{Acknowledgment}
% The preferred spelling of the word ``acknowledgment'' in America is without an ``e'' after the ``g.''
% Try to avoid the stilted expression, ``One of us (R. B. G.) thanks ...''
% Instead, try ``R.B.G. thanks ...''
% Put sponsor acknowledgments in the unnumbered footnote on the first page.

% \begin{thebibliography}{1}

% \bibitem{1}
% G.~Eason, B.~Noble, and I.~N.~Sneddon, ``On certain integrals of
% Lipschitz-Hankel type involving products of Bessel functions,''
% \emph{Phil. Trans. Roy. Soc. London,} vol. A247, pp. 529-551, April
% 1955.

% \bibitem{2}
% J.~Clerk~Maxwell, \emph{A Treatise on Electricity and Magnetism,}
% 3$^{\rm rd}$ ed., vol. 2. Oxford: Clarendon, 1892, pp.68-73.

% \bibitem{3}
% I.~S.~Jacobs and C.~P.~Bean, ``Fine particles, thin films and exchange
% anisotropy,'' in \emph{Magnetism,} vol. III, G.T. Rado and H. Suhl,
% Eds. New York: Academic, 1963, pp. 271-350.

% \bibitem{4}
% K.~Elissa, ``Title of paper if known,'' unpublished.

% \bibitem{5}
% R.~Nicole, ``Title of paper with only first word capitalized,''
% \emph{J. Name Stand. Abbrev.,} in press.

% \bibitem{6}
% Y.~Yorozu, M.~Hirano, K.~Oka, and Y.~Tagawa, ``Electron spectroscopy
% studies on magneto-optical media and plastic substrate interface,''
% \emph{APSIPA Transl. J. Magn. Japan,} vol. 2, pp. 740-741, August 1987
% [\emph{Digests 9$^{\rm th}$ Annual Conf. Magnetics Japan,} p. 301,
% 1982].

% \bibitem{7}
% M.~Young, \emph{The Technical Writer's Handbook.} Mill Valley, CA:
% University Science, 1989.

% \end{thebibliography}
\bibliographystyle{IEEEtran}
\bibliography{mybib}
\end{document}